\documentclass[prl,twocolumn,superscriptaddress]{revtex4}

\usepackage{graphicx}
\usepackage[normalem]{ulem} %need for strikethrough
\usepackage{color} %needed for coloring of \todo etc.

\begin{document}

\title{Optical vector network analysis of ultra-narrow transitions in $^{166}$Er$^{3+}$:$^7$LiYF$_4$}

\author{N.~Kukharchyk}
\affiliation{Experimentalphysik, Universit\"at des Saarlandes, D-66123 Saarbr\"{u}cken, Germany}

\author{D.~Sholokhov}
\affiliation{Experimentalphysik, Universit\"at des Saarlandes, D-66123 Saarbr\"{u}cken, Germany}

\author{O.~Morozov}
\affiliation{Kazan Federal University, 420008 Kazan, Russian Federation}

\author{S.~L.~Korableva}
\affiliation{Kazan Federal University, 420008 Kazan, Russian Federation}

\author{J.~H.~Cole}
\affiliation{Chemical and Quantum Physics, School of Science, RMIT University, Melbourne 3001, Australia}

\author{A.~A.~Kalachev}
\affiliation{Zavoisky Physical-Technical Institute, 420029 Kazan, Russian Federation}

\author{P.~A.~Bushev}
\affiliation{Experimentalphysik, Universit\"at des Saarlandes, D-66123 Saarbr\"{u}cken, Germany}

\date{\today}

\begin{abstract}
We present optical vector network analysis (OVNA) of an isotopically purified $^{166}$Er$^{3+}$:$^7$LiYF$_4$ crystal. The OVNA method is based on generation and detection of modulated optical sideband by using a radio-frequency vector network analyzer. This technique is widely used in the field of microwave photonics for the characterization of optical responses of optical devices such as filters and high-Q resonators. However, dense solid-state atomic ensembles induce a large phase shift on one of the optical sidebands which results in the appearance of extra features on the measured transmission response. We present a simple theoretical model which accurately describes the observed spectra and helps to reconstruct the absorption profile of a solid-state atomic ensemble as well as corresponding change of the refractive index in the vicinity of atomic resonances.  
\end{abstract}

\maketitle

Amplitude modulation (AM) spectroscopy represents one of the widely exploited modulation techniques in the rapidly advancing field of microwave photonics, which merges the worlds of radio-frequency and opto-electronics~\cite{Capmany2013}. Modulation methods are used to convert radio-frequency (RF) signals into the optical domain and vica versa in order to characterize optical components such as high-Q optical filters~\cite{Zhang2013,Li2017}, to perform signal processing and to filter-out undesired sidebands~\cite{Capmany2007}. The basic idea of the implemented technique is to use a RF vector network analyzer to modulate the optical signal with its own RF signal and detect its magnitude and phase delay by performing transmission measurement in a closed opto-electronic loop. In this context, generation of optical sidebands by precisely controlled RF/microwave signals opens up new possibility for quantum information processing with rare-earth (RE) doped solids. 
Particularly one can implement high-resolution spectroscopy, optical hole burning~\cite{Sellars2016}, electromagnetically induced transparency, state manipulation and Raman echoes by using very powerful methods of RF vector network analysis in a continuous as well as in a pulsed regime, such as those which have been developed for superconducting quantum circuits~\cite{Astafiev2010,Astafiev2011}. 

In the present work we perform crucial step towards the realization of quantum control on solid-state atomic ensembles in a closed opto-electronic loop. We demonstrate optical vector network analysis (OVNA) by using normal AM spectroscopy of ultra-narrow $^4I_{15/2} \leftrightarrow ^4I_{13/2}$ optical transition of isotopically purified $^{166}$Er$^{3+}$:$^7$LiYF$_4$ (Er:LYF) crystal. We also present a theory which accurately describes the observed modulation response, enabling the extraction of the absorption coefficient and change of refractive index in the vicinity of an atomic resonance.

\begin{figure}[htbp]
\centering
\includegraphics[width=0.8\linewidth]{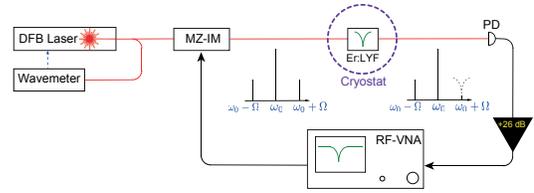}
\caption{Sketch of the experimental setup. MZ-IM stands for the Mach-Zehnder intensity modulator. PD is the high-speed InGaAs photodetector. RF-VNA is the radio-frequency vector network analyzer.}
\label{Setup}
\end{figure}

Isotopically purified, low-strain crystals doped with specific isotopes of rare-earth (RE) ions represent rather interesting materials for the implementation of optical quantum memory. Due to the ordered crystal structure, the inhomogeneous broadening of optical transitions of RE ions is mainly limited by nuclear spin fields of a host matrix~\cite{Popova1991,Macfarlane1992}. For instance, previous spectroscopic studies of $^7$LiYF$_4$ crystals doped with Er$^{3+}$ and Nd$^{3+}$ ions demonstrated optical FWHM inhomogeneous broadening in the order of $\Gamma/2\pi \sim 10-100$~MHz~\cite{Macfarlane1998,Gerasimov2016}. The presence of optical transitions of Er$^{3+}$ within the fibre transparency windows makes such solid-state media very attractive for realization of quantum memory protocols based on off-resonant Raman echoes, which promises to attain the highest quantum efficiency. The basic idea of such a protocol relies on using long-lived hyperfine transitions for the storage of optical photons. Isotopically purified crystals are the most suitable candidates for realization of off-resonant Raman memories because they offer direct optical addressing of their hyperfine states and large optical depth of $\alpha L > 10$. However, the accurate measurement of the high optical depth with conventional transmission spectroscopy is often very challenging due to finite resolution of oscilloscopes and difficulty of precise offset calibrations. Certainly, OVNA based on a single-sideband (SSB) detection method would represent the best option for the characterization of materials for quantum optical storage~\cite{Benito2004}. However, the implementation of SSB technique still requires deploying quite expensive and/or technically challenging solutions, such as dual-parallel Mach-Zehnder modulators~\cite{Zhu2014} or narrow-band tunable optical filters~\cite{Yao2012}. Contrary to that, we perform OVNA of a highly dispersive RE-doped solid by using conventional AM spectroscopy and explore its limitations.
 
\begin{figure}[htbp]
\centering
\includegraphics[width=0.8\linewidth]{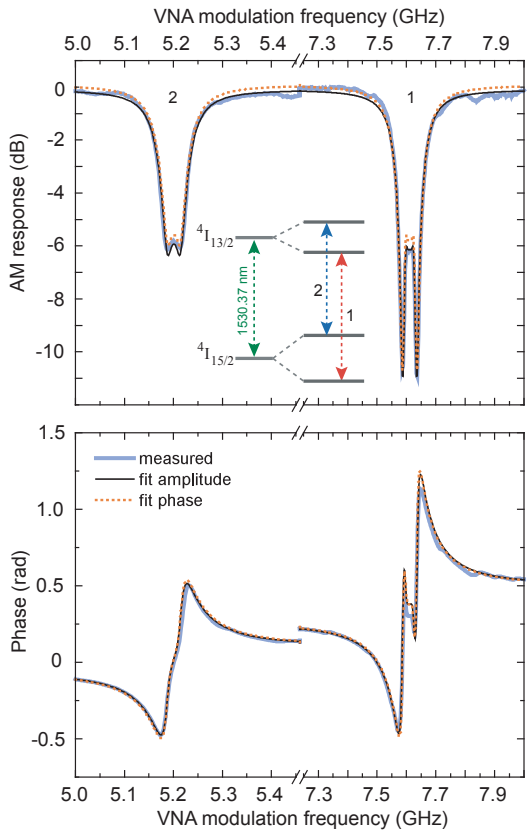}
\caption{Measured microwave transmission $|S_{21}(\Omega)|$ and phase delay arg~$( S_{21}(\Omega) )$ through the Er:LYF crystal.}
\label{VNA_response}
\end{figure}

The optical vector network analysis setup is sketched in Fig.~\ref{Setup}. The erbium-doped tunable fiber laser (NKT Photonics Adjustik E15) emits a continuous signal at the wavelength of $\lambda = 1530.37~$nm (vac) in the vicinity of the optical transitions of Er:LYF. The laser is tuned to $\omega_0/2\pi\approx195888.14~$GHz, which is 6 GHz away from the atomic transition at zero field, and its frequency is stabilized with the help of an optical wavemeter (High-Finess WS6-200). The optical carrier signal is modulated by applying a weak RF signal ($0$~dBm) from RF-VNA (Keysight PNA-L) to the Mach-Zehnder intensity modulator (MZ-IM), which is biased to its quad operating point. The spectral analysis of the generated optical signal shows that there is less than 1\% of the total light power contained in each of the generated optical sidebands. The laser light is focused into the sample, and the total light power at the sample is about 100~$\mu$W. 

The sample is cooled to 3.8~K. The external magnetic field is applied perpendicular to the crystal axis, and the polarization of the laser light is along the crystal axis $c$. The transmitted light is detected by a high-speed photodetector (Newport 1414A), which has a bandwidth of 25 GHz. The resulting microwave signal is amplified by a broadband 0.1-18 GHz, +26 dB microwave amplifier (Minicicrcuits ZVA-183S+) and is detected by the RF-VNA.

Our isotopically purified Er:LYF crystal has 0.0013\% doping concentration and a length of $L=0.5~$cm. This sample has been recently used in our previous experiments on studying the optical coherence at sub-Kelvin temperatures~\cite{Kukharchyk2017}. 

The measured microwave response ($|S_{21}|$, arg$(S_{21})$) of the crystal at the magnetic field of $B=0.2~$T is shown in Fig.~\ref{VNA_response}. The magnitude of the transmitted signal $\vert S_{21}(\Omega) \vert$ consists of two narrowband absorption components, namely 1 and 2, which are centered around $\Omega_1/2\pi=7.6$~MHz and $\Omega_2/2\pi=5.2~$GHz respectively. These components are associated with the optical transitions between Zeeman states with the same magnetic spin number $m_S$. Each spectral component reveals a double-peak feature. A similar feature is clearly recognizable in the phase response and is associated with a large phase delay acquired by the interacting optical sideband while it is swept across an atomic resonance. This auxiliary effect disappears for the transition 2 at higher fields, when the Zeeman level of optical ground state $^4I_{15/2}$ becomes depopulated, and optical density becomes small. Below, we outline our simple theoretical model which properly describes the observed effect.

The response of any dispersive medium is characterized by the complex susceptibility $\chi(\omega)=\chi`(\omega)+i\chi``(\omega)$, where its real and imaginary parts are connected via well-known Kramers-Kronig relations~\cite{Fabre_QO}. In the case of weak susceptibility, i.e. $\vert \chi`(\omega) \vert, \vert \chi``(\omega) \vert \ll 1$, the corresponding change of the refractive index and the absorption coefficient can be expressed via real and imaginary parts of $\chi(\omega)$ as $\Delta n (\omega) \approx \chi`(\omega)/2 $ and $\alpha(\omega) \approx \frac{\omega}{c} \chi``(\omega) $. By assuming a Lorentzian absorption profile, which is the case for ultra-narrow optical transitions in Er:LYF crystal~\cite{Kukharchyk2017}, we define the absorption coefficient as 
\begin{equation}
\alpha(\omega)=\alpha_0 \left( \frac{\Gamma^2}{ \Gamma^2 + 4 (\omega-\omega_r)^2}\right ),
\label{alpha}
\end{equation}
where $\alpha_0$ is the absorption coefficient of the medium at the atomic transition frequency $\omega_r$, and $\Gamma$ is the FWHM. The intensity absorption profile, i.e. the relative change of the intensity during propagation through the medium of length $L$, is determined by the Bugger law $A(\omega)=e^{-\alpha(\omega) L}$. The corresponding phase delay due to the dispersion is defined as $\phi(\omega)=\frac{\omega}{c} L \Delta n (\omega) =\alpha(\omega) L (\omega-\omega_r) / \Gamma$~\cite{Demtroder}. 

When a nearly-resonant laser wave propagates through an atomic ensemble it experiences both: absorption and phase delay (dispersion). In our experiment and thus in our model, only the blue-shifted optical sideband at $\omega_0+\Omega$ is assumed to interact with the ions, while the carrier at $\omega_0$ and the red-shifted sideband at $\omega_0-\Omega$ remain unperturbed. The beat between the modulated optical sidebands and the carrier is detected by the photodetector and results in the following RF signal at the modulation frequency of $\Omega$,
\begin{equation}
i_{RF}\propto  \cos (\Omega t) + A(\omega_0+\Omega) \cos (\Omega t + \phi(\omega_0+\Omega)).
\label{RF_beat}
\end{equation}
Two components of the RF signal are attributed to the beating of the carrier with red- and blue-shifted sidebands. The detection of the RF signal by VNA yields the microwave transmission magnitude and phase response which read as
\begin{eqnarray}
\vert S_{21}\vert^2 \propto 1 + A(\omega_0+\Omega)^2+2A(\omega_0+\Omega)\cos \phi(\omega_0+\Omega)  \\
arg(S_{21}) = \arctan\left( \frac{-A(\omega_0+\Omega)\sin \phi(\omega_0+\Omega)}{1+A(\omega_0+\Omega)\cos \phi(\omega_0+\Omega)}   \right)
\label{VNA_response_th}
\end{eqnarray}

The fit of the measured signal by the equations above is shown in Fig.(\ref{VNA_response}) with solid and dotted curves. At first, the magnitude response is fit by Eq.(3), and the phase response is fit by Eq.(4). Then, the parameters extracted from fit of the magnitude ($\alpha L$, $\omega_r$ and $\Gamma$) are used to simulate the phase response and visa versa. As it is seen from the graph, the measured data are in a very good agreement with the above presented theoretical model. Double peak structure appears in the absorption profile when the VNA response exceeds -6 dB.

\begin{figure}[htbp]
\centering
\includegraphics[width=0.8\linewidth]{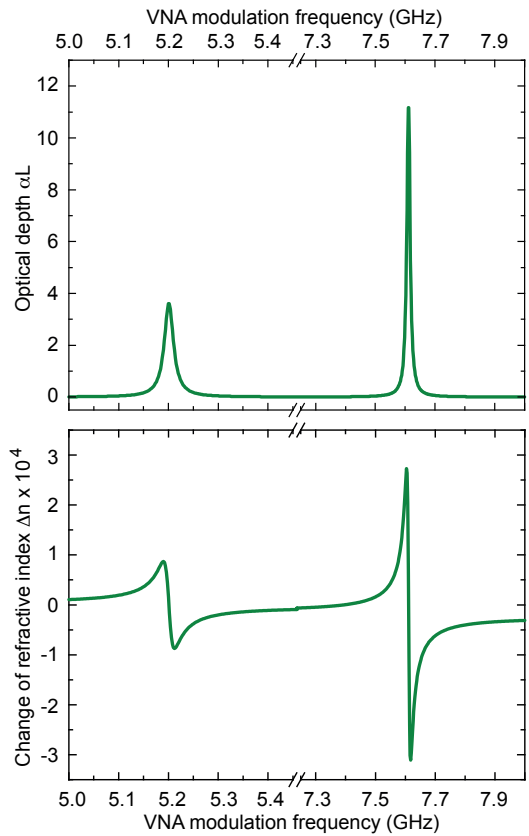}
\caption{Recovered optical depth $\alpha L$ and the change of the refractive index $\Delta n$ in the vicinity of atomic resonances by using OVNA.}
\label{Reconstruction}
\end{figure}

By using our theoretical model, we reconstruct the optical depth of the studied optical transitions and the corresponding change of the refractive index of a crystal around atomic resonance. The result of calculations is shown in Fig.~\ref{Reconstruction}. The maximal optical depth for transitions 1 and 2 is equal to 11.4 and 3.6 with the inhomogeneous broadening of $\Gamma_1/2\pi=13$~MHz and $\Gamma_2/2\pi=22$~MHz, respectively. The optical absorption coefficient for the transition 1 is $\alpha_1=22$~cm$^{-1}$. In a previous work optical depth as high as  $\alpha\simeq 40$~cm$^{-1}$ at $\Gamma/2\pi=16~$MHz was reported for a similar Er:LYF system at 0.005\% doping concentration~\cite{Thiel2011}. We note here that the deep freezing of the LYF crystal to sub-Kelvin temperatures pumps the population into the ground Zeeman state and results in similar values of $\alpha$ for the line 1 at magnetic field above $100~$mT~\cite{Kukharchyk2017}.

%The linewdith of optical transitions for our LYF system is determined by local magnetic field variations created by surrounding nuclear spins of the host matrix and reads as, see~\cite{Kukharchyk2017}:
%\begin{equation}
%\Gamma(B)= \Gamma_0+S_1 (B) \delta B + S_1(B) \alpha B, 
%\label{Linewdith}	
%\end{equation}
%where $\Gamma_0$ is the intrinsic inhomogeneous broadening, $S_1(B)$ is the first order curvature of the optical transition, $\delta B \simeq 1~$mT is the variation of local nuclear spin fields and $\alpha \simeq 3\cdot10^{-4}$ is the parameter describing inhomogeneity of the crystal and applied magnetic field. 
%
%The spectra evaluation for complete data set measured at different fields yields $\Gamma_0/2\pi=12~$MHz, which is up to the best of our knowledge represents the narrowest linewdith measured so far at telecom-C wavelength for the solid-state ensembles~\cite{Thiel2011,Gerasimov2016}. 

\begin{figure}[htbp]
\centering
\includegraphics[width=0.8\linewidth]{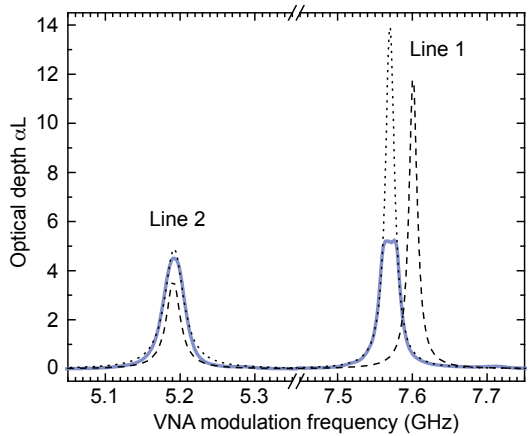}
\caption{Comparison between the optical depth measured by using the conventional transmission spectroscopy and OVNA. Solid line shows the conventional transmission spectrum and dotted line is the fit to Lorentzians. Dashed line shows the recovered optical depth measured by OVNA.}
\label{Comparison}
\end{figure}

It is interesting to compare the data measured with the presented method with a standard transmission spectroscopy. For this purpose, the modulation is turned off, and the laser frequency is swept by applying a triangular signal the piezo-transducer (PZT). The change of the laser wavelength is measured by optical wavemeter. The transmitted optical signal is detected by an amplified, slow InGaAs photodetector, and the detected signal is recorded by digital oscilloscope. The measured optical depth $\alpha L$ is shown in Fig.~\ref{Comparison} with a solid line. The absorption signal measured for the line 1 is clipped at $\alpha L \sim 5.5 $ due to the finite vertical resolution of the oscilloscope. The dotted line shows the fit of the data with a Lorentzian spectral shape excluding central truncated region, which yields $\alpha L $ to be equal 14 and 4.9 for line 1 and line 2 correspondingly. 

\begin{figure}[htbp]
\centering
\includegraphics[width=0.8\linewidth]{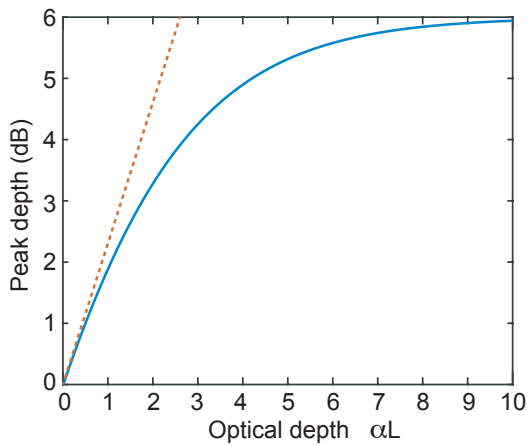}
\caption{Solid line: The measured depth of the absorption peak at the resonance frequency of the atomic transition $\omega_0+\Omega = \omega_r$ as a function of the optical depth of the sample $\alpha L$. Dotted line represents linear response of OVNA at $\alpha L \ll 1$.}
\label{dB_versus_alpha L}
\end{figure}

The dashed line in the Fig.~\ref{Comparison} shows the absorption reconstructed from AM data. The discrepancy in $\alpha L$ between the two data sets can be explained by the difficulty of a precise no-light-signal calibration in the conventional spectroscopy. We found that it can vary by few percents during the experiment. However, in the case of strong optical absorption, even such little offset variations can introduce a noticeable change into the $\alpha L$. The difference of the measured resonance frequency of about 20~MHz between the sets is due to the non-linear response of the PZT with respect to the applied voltage. In this experiment, the laser frequency is recorded while being swept. Then, the recorded signal is approximated by a 5-order polynomial function. The described procedure can result in $\sim 10~$MHz errors in the laser frequency determination. This discrepancy of the frequency scale also influences the linewidths of the transitions, which are found to be slightly higher for the conventional spectroscopy in comparison to the AM spectroscopy. Linewidths $\Gamma/2\pi$ measured by conventional spectroscopy are 14~MHz and 30~MHz, whereas by AM spectroscopy they are 13~MHz and 22~MHz for lines 1 and 2 respectively. 

The AM spectroscopy technique yields a non-linear response for large optical depths. To illustrate the limitations of the method we plot the absorption peak depth, measured by RF VNA at the atomic transition frequency $\omega_0+\Omega=\omega_r$, as a function of actual $\alpha L$ in Fig.~\ref{dB_versus_alpha L}. For a medium with $\alpha L \ll 1$ the quadratic magnitude term and phase shift in Eq.3 can be neglected, then it simplifies to $\vert S_{21}\vert^2(\Omega) \propto 1-\alpha(\omega_0+\Omega)L$. For the optical depth below 3, i.e. before the double-peak structure appears, it would be possible to convert the dB-signal directly into $\alpha L$ with a relatively simple polynomial function, which can be extracted from the curve on Fig.~\ref{dB_versus_alpha L}. We also note here, that double peak structure appears already at $\alpha L \simeq 4$. With increasing optical depth, the signal starts to significantly deviate from linear behavior and saturates at 6 dB, which corresponds to the complete removal one of the sidebands in the RF signal. At such high optical depth a fit with our model is required in order to extract the real parameters.

%For weakly absorbing transition with $\alpha L \ll 1$ the peak depth scales linearly and its magnitude in dB unit is equal to $\alpha L \log_{10} e $. In addition, the spectral shape of AM response is identical to the real one. For example, transitions between states having different $m_S$ have Gaussian shape due to the increased sensitivity to the local variations of magnetic field and that is detected by OVNA.

In conclusion, we presented optical vector network analysis of a record narrow (13~MHz) solid-state atomic ensembles with large optical density at telecom-C band. The auxiliary features observed on the AM response are attributed to the rapid phase change when one of the sidebands passes through the atomic resonance. We developed a simple theoretical model which accurately describes the measured modulation response of the medium. The model allows us to reconstruct the absorption profile and the refractive index change. We also compared the OVNA spectroscopy with conventional transmission spectroscopy and found reasonable agreement between the extracted parameters by using both methods. We also emphasize here, that using the OVNA can potentially allow for a more precise determination of Hamiltonian parameters of the ground and excited states of isotopically purified crystals with optically resolved hyperfine states. A pair of ultra-narrow resonant transitions with a tunable splitting allows implementing controllable optical delays using slow light~\cite{Camacho2006} and potentially may be useful for implementing quantum memory schemes involving refractive index control~\cite{Kalachev2012,Clark2012}. In addition, ultra-narrow solid state atomic ensembles are very attractive for the resonant filtering and conversion of optical signals~\cite{Shakhmuratov2017}. Overall, the presented experiment lays the foundation for the further integration of microwave and optical quantum systems which is required for the development of the quantum internet.  
%
%For the case of the moderate optical depths $\alpha L = 0.5-4$ the precise addressing of spin states of the rare-earth ions via microwave signals can be very helpful for the measuring of spin dynamics via hole burning and optically controlled spin echos~\cite{Sellars2016}. The implementation of SSB methods used in microwave photonics will allow to circumvent the limitation imposed by conventional AM spectroscopy and to work with materials having $\alpha L \gg 10$. 

\section{Funding Information}

DFG  the grant INST 256/415-1, BU 2510/2-1; RSF grant No. 14-12-00806; ARC Centre of Excellence
in Exciton Science (CE170100026).

\bigskip
\noindent

% Bibliography
\bibliography{AM_spectrum}

%\bibliographyfullrefs{AM_spectrum}

\end{document}